\documentclass{jetpl} \twocolumn
\usepackage{amsmath}
\usepackage{amsthm}
\usepackage{amsfonts}
\usepackage{amssymb}
\usepackage{graphicx}
\newcommand{\lb}[1]{\label{#1}}
\newcommand{\nn}{\nonumber}
\newcommand{\sbr}[1]{{\langle #1 \rangle}} 
\newcommand{\dbr}[1]{{\langle\!\langle #1 \rangle\!\rangle}} 
\newcommand{\beq}{\begin{equation}}
\newcommand{\ee}{\end{equation}}
\newcommand{\bea}{\begin{eqnarray*}}
\newcommand{\eea}{\end{eqnarray*}}

\input{epsf.tex}
\lat

\def\eqn{\begin{equation}}
\def\[{\begin{equation}}
\def\]{\end{equation}}

\def\exp{{\rm exp}}

\def\sin{{\rm sin}}

\title{On scaling fields in $Z_N$ Ising models}

\rtitle{On scaling fields in $Z_N$ [Ising] models}

\sodtitle{On scaling fields in $Z_N$ models}

\author{V.~A.~Fateev$^{a,b}$, V.~V.~Postnikov$^c$, Y.~P.~Pugai$^a$
}

\rauthor{V.~A.~Fateev, V.~V.~Postnikov, Y.~P.~Pugai}

\sodauthor{V.~A.~Fateev, V.~V.~Postnikov, Y.~P.~Pugai}

\address{$^a$L.~D.~Landau Institute for Theoretical Physics RAS,
117940 Moscow, Russia}

\address{$^b$ Laboratoire de Physique
Th\'eorique
et Astroparticules, Universit\'e Montpellier II,  34095 Montpellier, France}

\address{$^c$ Sochi's branch of People Friendship University, Sochi, Russia}
\dates{\today}{*}
\abstract{We study the space of scaling fields in the $Z_N$
symmetric models with the factorized scattering and propose simplest
algebraic relations between form factors induced by the action of
deformed parafermionic currents. The construction gives a new free
field representation for form factors of perturbed
Virasoro algebra primary
fields, which are parafermionic algebra descendants.
We find exact vacuum expectation values of
physically important fields and study correlation functions of order
and disorder fields in the form factor and CFT perturbation
approaches.}


\begin{document}
 \maketitle


{\it 1. Introduction}

\par\noindent
The calculation of exact correlation functions is one of the most complicated
and interesting problems in the two dimensional integrable quantum
field theory.

Remarkable solutions of this problem were found for the free field
models including the Ising model without a magnetic field and for
the conformal field theories (CFT) \cite{BPZ,DotsFat,ZaFa85}. One of
the difficulties in investigating massive integrable field theories
with a non-trivial interaction is related with the fact that the
space of scaling fields and, in particular, the space of different
solutions of form factor equations, has no clear and simple
description in terms of dynamical symmetry algebras like the free
fermion algebra in the Ising model or Virasoro algebra in CFT.
Existence of such infinite-dimensional algebras drastically
simplifies the problem and allows applications of powerful algebraic
methods. For instance in CFTs the Ward identities generated by the
holomorphic stress energy tensor $T(z)$ permit to express the
correlation functions of Virasoro descendant fields in terms of the
correlators of the primary fields and to develop the free field
approach for their calculation in the minimal models of CFT
\cite{DotsFat}.

In massive integrable theories the situation with descendant fields is
more difficult. Perturbing away from the conformal invariant point
\cite{Zam} one expects that the space of scaling fields is in a
one-to-one correspondence with that of conformal fields. This
statement was supported by several results on counting solutions of
form factor equations \cite{KaWe78,smirnovbook} in massive
integrable field theories which demonstrate appearance of conformal
algebra characters even off criticality \cite{Jim}. This phenomena
may indicate existence of some deformed conformal algebra action in
the space of form factors which would allow to extract useful
information on descendants from knowledge on primaries (see also
recent works in this direction \cite{Del}).

In this work we study the structure of the space of form factors of
scaling fields in the parafermionic CFT with the central charge
\[
\label{Charge} c={2(N-1)}/({N+2})\,,\quad  N=2,3,\cdots \,,\]
perturbed by the first thermal operator \cite{KoSw,Fat91}. In the
free field approach \cite{Luk95,ZF} to form factors we develop
further the idea of the paper \cite{Smir} on relations between
different form factors of the theory. Namely, we introduce actions
of simplest parafermionic modes in the space of form factors by
exploiting the deformed parafermionic symmetry \cite{ABFII} of the
underlying lattice model \cite{ABF}. We check consistency of our
construction by numerical studying of correlation functions of the
theory by using a combination of form factor and conformal
perturbation theory approaches \cite{Zam91}. To do this we also
compute new exact VEVs of physically important fields.

\vspace{0.2cm} {\it 2. Space of states of parafermionic CFT
\cite{ZaFa85}}

\par\noindent
Conformal field theories with the parafermionic symmetry describe
self-dual critical points of $Z_N$ ($N=2,3,\ldots$) generalizations
\cite{ZaFa85,ABF} of the Ising model, the last one corresponds to
the $N=2$ case. In a continuum limit order parameters $\sigma_k(x)$
($k=1,\ldots,N-1$) determining long range correlations of spins have
anomalous dimensions
\[\label{SpinDim}
2d_k=\frac{k(N-k)}{N(N+2)}\,.\]
Their $Z_N$ charge is equal to $k$, i.e., under the action of the
$Z_N$ symmetry they transform as
\[ \sigma_{k}\rightarrow\omega^{kn}\sigma_{k}, \quad n\in Z\,.
\label{SigmZ}
\]
$Z_N$ models allow Kramers-Wannier symmetry and the theory also
contains the disorder parameters $\mu_{k}$ ($k=1,\ldots,N-1$) with
the same dimensions $2d_k$ at the self-dual point. Operators
$\mu_{k}$ transform under the action of the dual $\tilde{Z}_N$
symmetry as
\[
\quad\mu_{k}\rightarrow \omega^{kn}\mu_{k},\quad n\in Z\,. \]
All other fields naturally separating into families with the fixed
$Z_{N}\times\widetilde{Z}_{N}$ charges $(k,l)$ behave as
$$
\phi\rightarrow\omega^{kn+ln'}\phi, \quad n,n'\in Z\,.
$$

The parafermions $\psi_{k}$ and $\overline{\psi}_{k}$ generalizing
usual fermions appear in the OPE of the order and disorder fields
\begin{eqnarray}
&&\sigma_{k}(z,\overline{z})\mu_{k}(0,0)  =|z|^{-4d_{k}}z^{\Delta_{k}}%
[\ \psi_{k}(0)+\cdots]\,, \cr &&
\sigma_{k}(z,\overline{z})\mu_{k}^{+}(0,0)  =|z|^{-4d_{k}}\overline
{z}^{\Delta_{k}}[\ \overline{\psi}_{k}(0)+\cdots]\,. \end{eqnarray}
These currents are holomorphic and generate the infinitely
dimensional symmetry due to the conservation laws
\[
\partial_{\overline{z}}\psi_{k}=0\,,\qquad
\partial_{z}\bar{\psi}_{k}=0\,.
\label{ConsLaw}
\]
We concentrate on the simplest solution of the associativity
condition for the operator algebra of currents
\[
\psi_{k}(z)\psi_{l}(0)=z^{\Delta_{k+l}-\Delta_{k}-\Delta_{l}}\left[
\psi_{k+l}+...\right]\,, \]
which corresponds to the conformal
field theory with the central charge (\ref{Charge}) and the
conformal dimensions
\[ \Delta_{k}=\frac{k(N-k)}{N}\,. \]
The fields $\psi_{1}\equiv\psi$
($\overline{\psi}_{1}\equiv\overline{\psi}$) \ are the basic ones in the
pa\-ra\-fer\-mi\-onic algebra. It will be convenient for us to
consider as well conjugate currents $\psi^\dagger\equiv \psi_{N-1}$
($\bar{\psi}^\dagger\equiv \bar{\psi}_{N-1}$).

In the conformal models the space of states splits naturally into a
direct sum of subspaces with the specified
$Z_{N}\times\widetilde{Z}_{N}$  charge $(k,l)$
\[ {\{F\}}=\oplus   \ {\{F\}}_{\left[ m,\overline{m}\right] } ,\quad
N\geq m,\overline{m}\geq1-N_{{}}\,,%
\]
where $\left[ m,\overline{m} \right] =\left[k+l,k-l\right]
,m+\overline{m} \in2Z$.
%
%
In these notations parafermionic currents and order-disorder fields
belong to the following subspaces
\begin{eqnarray} &&\ \psi\in {\{F\}}_{\left[  2,0\right] }\,, \quad \
\psi^\dagger\in {\{F\}}_{\left[ -2,0\right] }\,,\label{Spaces}\\ &&\
\overline{\psi}\in {\{F\}}_{\left[ 0,2\right] }\,, \quad \
\overline{\psi}^\dagger\in {\{F\}}_{\left[ 0,-2\right] }\,, \cr &&
\sigma_{k}\in {\{F\}}_{\left[ k,k\right] }\,, \ \quad \mu_{k}\in
{\{F\}}_{\left[ k,-k\right]  }\nn \,.\end{eqnarray}

Conformal fields are conveniently classified according to the
representations of the pa\-ra\-fer\-mi\-o\-nic algebra. The action
of the parafermionic generators $A_{\nu}$ ($A_{\nu}^\dagger$) is
defined by the OPE
\begin{eqnarray} && \psi(z)\phi_{[m,\bar{m}]}=\sum
z^{-\frac{m}{N}+n-1}A_{\frac{1+m}{N}-n}\ \phi_{[m,\bar{m}]}\,, \label{PFactionCFT}\\
\vspace{0.2cm} && \psi^\dagger(z)\phi_{[m,\bar{m}]}=\sum
z^{\frac{m}{N}+n-1}A^\dagger_{\frac{1-m}{N}-n}\
\phi_{[m,\bar{m}]}\,.\nn
\end{eqnarray}
Notice that, if $\phi_{[m,\bar{m}]}\in F_{[m,\bar{m}]}$ has the
conformal dimensions $(d,\bar{d})$ then the conformal dimensions of
fields
\begin{eqnarray}
&& A_{\nu}\phi_{[m,\bar{m}]}\in \{F\}_{[m+2,\bar{m}]} \,,\\ &&
A^\dagger_{\nu}\phi_{[m,\bar{m}]}\in \{F\}_{[m-2,\bar{m}]} \nn\,,
\end{eqnarray}
are $(d-\nu,\bar{d})$.

Basic fields $\mu_{k}$ $\left(\sigma_{k} \right) $ are the primaries
of the pa\-ra\-fer\-mi\-onic algebra. For instance, the following
equations hold for $n\geq  0$
\begin{eqnarray} && A_{\frac{1+k}{N}+n}\ \mu_{k}=A^\dagger_{\frac{1-k}{N}+n+1}\
\mu_{k}=0\,,\label{HighestW}\\ && \bar{A}_{\frac{1-k}{N}+n+1}\
\mu_{k}=\bar{A}^\dagger_{\frac{1+k}{N}+n}\ \mu_{k}=0\,.\nn
\end{eqnarray}
%
%
%
All other fields of the model are obtained by the action of the
currents $\psi,\bar{\psi}$ on $\mu_k \ (\sigma_k)$. Thus, the space
of states of the conformal field theory decomposes into a direct sum
of irreducible representations of the parafermionic algebra
\[
\{F\}=\oplus_{k=0}^{N-1}\ [\mu_k]_A\,,
\]
with the disorder operators $\mu_k$ playing the role of the highest
weight fields.

However, from the viewpoint of Virasoro algebra, each of the spaces
$[\mu_k]_A$ expands into a direct sum of representations
with the highest weight fields $\phi_{m,\bar{m}}^{k}$. These Virasoro
primaries are, in general, descendants in the parafermionic algebra
representation. Indeed, let us denote as $(\psi^\dagger)^{l}\mu_{k}$
the field with the minimal conformal dimension that can be obtained
by $l-$times application of the parafermionic generators
$\psi^\dagger$ to $\mu_{k}$ (and respectively for $\bar{\psi}$)
\begin{eqnarray}
(\psi^\dagger)^{l}\mu_{k}=
A^\dagger_{\frac{2l-1-k}{N}}A^\dagger_{\frac{2l-3-k}{N}}\cdots
A^\dagger_{\frac{1-k}{N}}\mu _{k}\,. \end{eqnarray}
Its conformal dimensions are easily computed from (\ref{SpinDim}).
Then, up to a normalization, the following relations take place
\begin{eqnarray}
\label{Prim}
&&\phi_{k-2l,-k+2\bar{l}}^{k}=(\psi^{+})^{l}(\overline{\psi})^{\bar{l}%
}\mu_{k}\,,\quad  l,\bar{l}=0,1,...,k\,,\\ &&
\phi_{k+2l,-k-2\bar{l}}^{k}=(\psi)^{l}(\overline{\psi^\dagger})^{\bar{l}}
\mu_{k}\,, \quad l,\bar{l}=0,1,...,N-k\,.\nn\end{eqnarray}
%
For example, the physically important thermal operators
$\varepsilon_{k}=\phi_{0,0}^{2k}$ are among these primaries. Recall
that $\varepsilon_{k}$ are local with respect to all fields and have
the conformal dimensions
\[ D_{k}={k(k+1)}/{(N+2)}.\]


\vspace{0.2cm} {\it 3. Form factors for perturbed primary fields}
\par\noindent
We consider the perturbation of the parafermionic conformal field
theory by the first energy operator
\[
\mathcal{A}=\mathcal{A}_{CFT}+\lambda\int d^{2}x\
\varepsilon_{1}(x)\,.
\]
The resulting massive theory is again integrable and $Z_N$
symmetric. Depending on the sign of $\lambda$ the system is in the
ordered or disordered phase. We fix $\lambda>0$ phase where the
$\tilde{Z}_N$ symmetry is destroyed and vacuum expectation values of
disorder operators are non-zero.

Alternatively, this theory can be described also as a model with
factorized scattering of $Z_N$ charged particles \cite{KoSw}. The
particles $a\in \{1,\ldots, N-1\}$ in the $Z_N$ $(N=2,3,4,\ldots)$
symmetric models have masses
\[ M_a=M\frac{\sin(\pi a /N)}{\sin(\pi /N)}\,. \]
Note, that the antiparticle ${a}^\dagger$ is identified with the
particle $N-a$. The scattering matrix of the lightest particles
$a=1$ has the form
\begin{eqnarray}
S_{11}(\beta)=
\frac{\sinh(\frac{\beta}{2}+\frac{i\pi}{N})}
{\sinh(\frac{\beta}{2}-\frac{i\pi}{N})}\,.
\end{eqnarray}
The S matrices for higher particles are also diagonal and can be
extracted from $S_{11}$ according to a standard bootstrap
prescription \cite{KoSw}.

We follow the algebraic approach \cite{Luk95,ZF,ABFII,VEV} to the
form factors originated from the analysis of lattice ABF model
\cite{ABF}. This lattice model falls into the same universality
class as the $Z_N$ Ising model. In the corner transfer matrix
approach its dynamical symmetry is a deformation of a parafermionic
symmetry. The deformed currents $\Psi(\alpha)$, as well as deformed
primaries $\Phi_{m}^{k}(\alpha)$ (which are primary for deformed
Virasoro algebra \cite{ZF}) act in the corner transfer matrix space
of states. Note, that there is an essential difference between the
symmetries in off-critical situation and in the CFT. Thus, there
exists only one chiral algebra, the parameter $\alpha$ is related
with the spectral parameter but not with the distance, etc. However,
from the mathematical viewpoint, the analysis of ABF models,
including fusions rules, $Z_N$ symmetry, the structure of
irreducible representations etc., is very similar to the conformal
situation (\ref{SigmZ})-(\ref{Spaces}). Respectively, as in CFT, the
problem of the calculation of physical quantities reduces to
studying of operators acting in the direct sum of representations of
the deformed parafermionic algebra.

In particular, the form factors are matrix elements of local
operators in the basis of asymptotic states formed by operators
diagonalizing the Hamiltonian. A peculiar feature of the $Z_N$
models is that the Hamiltonian is diagonalized by the symmetry
generators $\Psi$ ($\Psi^{\dagger}$) themselves. At the scaling
limit the currents become particle creation operators
${{\mathcal{B}}}$ (${{\mathcal{B}}}^{\dagger}$) which in the free
fields approach \cite{Luk95} can be described as following.
\begin{eqnarray} &&\mathcal{B}(\beta) =\frac{e^{-\frac{\beta}{N}P_2}}{
\sqrt{2\sin \frac{\pi}{N} } } \sum_{a=\pm 1}a\,e^{\frac{ \pi
i}{2N}(P_1+P_2+1)a} \mathcal{Z}_{a}(\beta),
\lb{expo1}\\
&&\mathcal{B}^\dagger(\beta) =-\frac{e^{\frac{\beta}{N}P_2}}{
\sqrt{2\sin \frac{\pi}{N} } }\quad \sum_{a=\pm 1}a\,e^{\frac{ \pi
i}{2N}(-P_1+P_2-1)a} \mathcal{Z}^\dagger_{a}(\beta). \nonumber
\end{eqnarray}
These linear combinations contain the exponentials of free bosons
$\mathcal{Z}_{\pm}(\beta)$ and $\mathcal{Z}^\dagger_{\pm}(\beta)$
and the "zero modes" operators $P_{1,2}$. The operators
$\mathcal{Z}_{a}(\beta)$ and $\mathcal{Z}_{a}^{\dagger}(\beta)$ are
assumed to satisfy the Wick theorem
\begin{eqnarray} &&
\dbr{\mathcal{Z}_{a_1}(\beta_1)\cdots\mathcal{Z}_{a_n}(\beta_n)
\mathcal{Z}_{b_1}^\dagger(\beta_1')\cdots\mathcal{Z}_{b_m}^\dagger(\beta_m')
  }
\label{Wick} \\
  &&=
\prod_{i<j}\dbr{\mathcal{Z}_{a_i}(\beta_i)\mathcal{Z}_{a_j}(\beta_j)}
\prod_{i<j}\dbr{\mathcal{Z}_{b_i}^\dagger(\beta_i')\mathcal{Z}_{b_j}^\dagger(\beta_j')}\times
\nn\\
  &&\quad
\prod_{
i,j}\dbr{\mathcal{Z}_{a_j}(\beta_j)\mathcal{Z}_{b_i}^\dagger(\beta_i')}\,,
\nn\end{eqnarray}
with the following contraction rules (where
$\beta=\beta_1-\beta_2$):
\begin{eqnarray} && \dbr{\mathcal{Z}_{a}(\beta_1)\mathcal{Z}_{b}(\beta_2)}
=\dbr{\mathcal{Z}^\dagger_{-a}(\beta_1)\mathcal{Z}^\dagger_{-b}(\beta_2)}
\\
&& \qquad= \zeta(\beta)
\frac{\sinh\bigl(\frac{\beta}{2}+\frac{i\pi}{2N}(a-b)\bigr)}
{\sinh\frac{\beta}{2}}\,, \cr &&
\dbr{\mathcal{Z}_{a}(\beta_1)\mathcal{Z}^\dagger_{b}(\beta_2)}
=\dbr{\mathcal{Z}^\dagger_{-a}(\beta_1)\mathcal{Z}_{-b}(\beta_2)}
\cr && \qquad =\zeta^{\dagger}(\beta)
\cosh\bigl(\frac{\beta}{2}-\frac{i\pi}{2 N}(a+b)\bigr)\,.\nonumber
\end{eqnarray}
The functions $\zeta(\beta)$, $\zeta^{\dagger}(\beta)$ and their
basic properties are given in the appendix A.

The perturbed primary fields $\phi^{k}_{m,-m}$ corresponds to the
projectors to the states $\langle\langle \Phi_{-m}^k|$ and
$|\Phi_{m}^k\rangle\rangle$. We treat the last ones as states
created by the deformed operators $\Phi^{k}_{\pm m}$ from the vacuum
vector \cite{VEV}
\begin{eqnarray}
\label{BrKt} |\Phi_{m}^k\rangle\rangle &&=\lim_{\alpha\rightarrow
\infty}\Phi^k_m(\alpha)|0\rangle\rangle
\,,\\
\langle\langle \Phi_{-m}^k|&&=\lim_{\alpha\rightarrow
-\infty}\langle\langle 0|\Phi_{-m}^{k}(\alpha) \,.\nn\end{eqnarray}
Let us note that these bra and ket vectors in the corner transfer
matrix picture correspond to the Virasoro algebra states created by
left and right chiral parts of fields $\phi_{m,-m}^k$. The action of
the zero modes on the vectors (\ref{BrKt}) reads as
\begin{eqnarray} &&P_1|\Phi_{m}^k\rangle\rangle=k|\Phi_{m}^k\rangle\rangle,\quad
P_2|\Phi_{m}^k\rangle\rangle=-m|\Phi_{m}^k\rangle\rangle
\,,\end{eqnarray}
and we assume that the operators $P_{j}$ commute with operators
$\mathcal{Z}_{a}(\beta)$, $\mathcal{Z}_{a}^{\dagger}(\beta)$. The
normalization of vectors $|\Phi_{m}^k\rangle\rangle$ is chosen
according to the prescription \cite{VEV}
\[
\dbr{\Phi_{-m}^k|\Phi_{m}^k}=\sbr{\phi^{k}_{m,-m}}\,,
\label{VEVLat}\]
which would correspond to
the conformal normalization of scaling fields
\[
\langle \phi(x) \phi(0) \rangle = {|x|^{-4\Delta_\phi}}, \ \ \
x\rightarrow 0\,. \label{ConfNorm}\]
Then the form factors of the Virasoro algebra primary fields $
{\phi}^{k}_{m,-m}$, which are local with respect to the fields
$\mu_k$, can be compactly written as following
\begin{multline}
  \lb{energy2}
\langle {\phi}^{k}_{m,-m}|\{\beta\},\{\beta'\}\rangle_{(n,n)}=\\
=
 \dbr{\Phi_{-m}^k|\prod^n\mathcal{B}(\beta_j)
\prod^n\mathcal{B}^\dagger(\beta_j')|\Phi_{m}^k}\,.
\end{multline}
We implied here a shorthand notation for the state with $n'$
lightest particles 1 and $n$ antiparticles $1^\dagger$
\bea && |\{\beta\},\{\beta'\}\rangle_{(n,n')}\equiv
|\beta_1,\ldots,\beta_n,\beta_1',\ldots,\beta_{n'}'\rangle_{
1^\dagger\cdots 1^\dagger,1\cdots 1} \,\eea
The symbols of ordered products of particle creation operators stand
for
\bea
\prod^n\mathcal{B}(\beta_j)=\mathcal{B}(\beta_1)\cdots
  \mathcal{B}(\beta_n)\,.
\eea
The equation (\ref{energy2}) is a natural generalization of the
thermal operators form factors \cite{ABFII} to the $m\neq 0$ case.
With the definitions (\ref{expo1})-(\ref{Wick}) it has the
conventional form
\begin{multline}
  \lb{energy3}
\langle {\phi}^{k}_{m,-m}|\{\beta\},\{\beta'\}\rangle_{(n,n)}
=\frac{(-1)^n}{(2\ \sin\frac{\pi}{N})^n}e^{\frac{m}{N}\sum
(\beta_j-\beta_j')}\times\\
\sum_{\{a_j,b_j\}}\prod_{j}a_jb_je^{\frac{i\pi
}{2N}((k+1)(a_j-b_j)-m(a_j+b_j))}\times
\\
 \dbr{\mathcal{Z}_{a_1}(\beta_1)
\cdots
\mathcal{Z}_{b_n}^\dagger(\beta_n')}\dbr{\Phi_{-m}^k|\Phi_{m}^k}\,.
\end{multline}

The remarkable property of our theory is that form factors of
perturbed Virasoro primaries $ {\phi}^{k}_{m,\bar{m}}$ including
(\ref{energy2}) can be obtained form factors of fields $\mu_k$ by
the equation analogous to (\ref{Prim}). Indeed, for large $\alpha$
the operators $\Phi^{k}_{\pm m}(\alpha)$ from (\ref{BrKt}) behave in
the conformal limit, roughly speaking, as chiral parts of primaries
$\phi_{m,-m}^k(e^{-\alpha},e^{-\bar{\alpha}})$. Respectively, the
operators $\Psi(\alpha)$ in CFT limit become the parafermionic
currents $\psi(e^{-\alpha})$. Basing on this correspondence, we
introduce the analogs of rules (\ref{PFactionCFT})-(\ref{Prim}) for
the action of the simplest Fourier modes of deformed currents $\Psi$
($\Psi^{\dagger}$) as
\begin{eqnarray}
|\Phi_{m+2}^k\rangle\rangle
=&&|A_{\frac{1+m}{N}}\Phi^{k}_{m}\rangle\rangle =\lim_{\alpha\to
\infty}
e^{-\frac{1+m}{N}\alpha}\Psi(\alpha)|\Phi_{m}^k\rangle\rangle\,, \cr
|\Phi_{m-2}^k\rangle\rangle=&&|{A}^\dagger_{\frac{{1-m}}{N}}\Phi^k_{m}\rangle\rangle
=\lim_{{\alpha}\to \infty}
e^{-\frac{1-{m}}{N}{\alpha}}\Psi^\dagger({\alpha})
|\Phi_{m}^k\rangle\rangle\,, \cr
\langle\langle \Phi_{\bar{m}+2}^k| =&&
\langle\langle\bar{A}_{\frac{1+\bar{m}}{N}}\Phi^k_{\bar{m}}|
=\lim_{{\alpha}\to -\infty} e^{\frac{1+\bar{m}}{N}{\alpha}}
\langle\langle\Phi_{\bar{m}}^k|\Psi({{\alpha}}) \,, \cr
\langle\langle\Phi_{\bar{m}-2}^k|=&&
\langle\langle\bar{A}^\dagger_{\frac{1-\bar{m}}{N}}\Phi^k_{\bar{m}}|
=\lim_{{\alpha}\to -\infty} e^{\frac{1-\bar{m}}{N}{\alpha}}
\langle\langle\Phi_{\bar{m}}^k|{\Psi}^\dagger({{\alpha}})\,.
\nonumber\end{eqnarray}
Since the operators $\mathcal{B}$ ($\mathcal{B}^\dagger$) are,
essentially, the scaling limits of the deformed pa\-ra\-fer\-mi\-ons
$\Psi$ ($\Psi^{\dagger}$), we put
\begin{eqnarray}
\label{PFOne}&&\langle
A_{\frac{1+m}{N}}\phi^k_{m,\bar{m}}|\{\beta\},\{\beta'\}\rangle_{(n,n')}=\\
&&\quad\lim_{\alpha\to \infty}
e^{-\frac{1+m}{N}\alpha}\dbr{\Phi_{\bar{m}}^k|
\prod^{n}\mathcal{B}(\beta_j)
  \prod^{n'}\mathcal{B}^\dagger(\beta'_j)\ \mathcal{B}(\alpha) |\Phi_{m}^k}
\,,\nonumber
\end{eqnarray}
\begin{eqnarray}
&&\langle\bar{A}_{\frac{1+\bar{m}}{N}}\phi^k_{m,\bar{m}}|\{\beta\},\{\beta\}\rangle_{(n,n')}=\\
&&\quad\lim_{\alpha\to -\infty}
e^{\frac{1+\bar{m}}{N}\alpha}\dbr{\Phi_{\bar{m}}^k|\mathcal{B}(\alpha)
 \prod^{n}\mathcal{B}(\beta_j)
  \prod^{n'}\mathcal{B}^\dagger(\beta'_{j})|\Phi_{m}^k}
\,.\nonumber
\end{eqnarray}
Starting from (\ref{energy2}) we have the $Z_N$ neutrality condition
$2(n'-n)=m+\bar{m}+2$. For Fourier modes $A^\dagger$, we obtain
\begin{eqnarray} &&\langle
A^\dagger_{\frac{1-m}{N}}\phi^k_{m,\bar{m}}|\{\beta\},\{\beta'\}\rangle_{(n,n')}=\\
&&\quad \lim_{\alpha\to \infty} e^{-\frac{1-m}{N}\alpha}\dbr{
\Phi_{\bar{m}}^k|
 \prod^{n}\mathcal{B}(\beta_j)
  \prod^{n'}\mathcal{B}^\dagger(\beta'_j)\ \mathcal{B}^\dagger(\alpha)|\Phi_{m}^k}
\,,\nonumber
\end{eqnarray}
\begin{eqnarray} && \label{PFTwo}
\langle\bar{A}^\dagger_{\frac{1-\bar{m}}{N}}\phi^k_{m,\bar{m}}|\{\beta\},\{\beta\}\rangle_{(n,n')}=\\
&& \quad\lim_{\alpha\to -\infty}
e^{\frac{1-\bar{m}}{N}\alpha}\dbr{\Phi_{\bar{m}}^k|
\mathcal{B}^\dagger(\alpha) \prod^{n}\mathcal{B}(\beta_j)
  \prod^{n'}\mathcal{B}^\dagger(\beta'_{j})|\Phi_{m}^k}
\,,\nonumber \end{eqnarray} with the selection rule $\
2(n'-n)=m+\bar{m}-2$.

These actions provide maps between solutions of form factor
equations for fields with the different $Z_N$ charges. Note, that
the equation (\ref{Prim}) holds for corresponding multi-particles
form factors, as it can be directly verified from (\ref{energy2})
and (\ref{PFOne}-\ref{PFTwo}). In this way we extend free field
formulae for primaries $\phi^{k}_{m,\bar{m}}$.


\vspace{0.2cm} {\it 4. Short distance expansions and VEVs}
\par\noindent
As an application for new form factor formulae (\ref{energy2}),
(\ref{PFOne}-\ref{PFTwo}) and for checking the consistency of the
proposed parafermionic actions we analyze the correlation functions
\[ G_{+}(x)=\langle\sigma_{1}(x)\sigma_{1}^{+}(0)\rangle\,,\ \
G_{-}(x)=\langle\mu_{1}(x)\mu_{1}^{+}(0)\rangle\,, \]
computing it in the form factor approach (long distance expansion)
from one side and in the
conformal perturbation theory method (short distance expansion)
\[
G_{\pm}(x)=r^{-4d_{1}}\sum C^{A_{l}}_\pm (r)\langle
A_{l}(0)\rangle\,, \ \  r=|x|\,,\label{ShortA}
\]
from the other side \cite{Zam91}. The structure functions
$C_{\pm}^{A_{l}}(r)$ above allow expansions into perturbation series
\bea C_{\pm}^{A_{l}}(r)=r^{2\Delta_{A_{l}}}\left(
C_{\pm,(0)}^{A_{l}}+ \lambda r^{2(1-D_1)}
C_{\pm,(1)}^{A_{l}}+\cdots\right)\,, \eea
where the coefficients $C_{\pm,(1)}^{A_{l}}$ can be expressed
through the integrals of correlation functions in CFT.

In the first order approximation the main contributions to
(\ref{ShortA}) come from the primaries
$I,\varepsilon_{1},\varepsilon_{2}$ and the W-algebra descendant field
$E_{1}=W_{-1}^{(3)}\overline{W}_{-1} ^{(3)}\varepsilon_{1}$
\begin{eqnarray}
&&C_{\pm}^{I}=1\pm\frac{\lambda\pi
r^{2(1-2u)}(\gamma(u)\gamma(3u))^{1/2}\gamma
(4u)}{2(1-4u)^{2}\gamma^{3}(2u)}%
\,, \label{ShortB}\\
&&C_{\pm}^{\varepsilon_{1}}=\mp\frac{r^{4u}(\gamma(u)\gamma(3u))^{1/2}}
{2\gamma(2u)}+\frac{\lambda\pi r^{2}u^{2}\gamma(4u)}
{4\gamma^{4}(2u)\gamma
^{4}(1-u)}%
\,,\cr &&C_{\pm}^{\varepsilon_{2}}=\mp r^{12u}\frac{\lambda\pi
r^{2(1-2u)}u^{2}\gamma
^{3}(u)(\gamma(5u))^{1/2}}{12(1+2u)^{2}\gamma^{2}(2u)(\gamma(3u))^{1/2}}%
\,,\cr &&C_{\pm}^{E_{1}}=-r^{2+4u}
\frac{u(1-4u)(\gamma(u)\gamma(3u))^{1/2}}{(1+2u)(1-2u)\gamma
(2u)}^{{}}\,.\nn
\end{eqnarray}
Here and later we denote
\[\gamma(z)=\frac{\Gamma(z)}{\Gamma(1-z)}\,, \quad
u=\frac{1}{N+2}\,.\]
Let us recall that the mass $M$ of the lightest particle and the
coupling constant $\lambda$ can be expressed via each other. The
exact relation between them in the conformal normalization
$\langle\varepsilon_{1}(x)\varepsilon_{1}(0)\rangle$
$=|x|^{-4D_{1}}$ is known \cite{Fat94} to have a form

\begin{eqnarray}
&&(2\pi\lambda)^{2}=\kappa^{4(1-2u)}\gamma(u)\gamma(3u)\,,\\ &&\cr
&&\quad \kappa =M\frac{\Gamma(2/N)\Gamma(1-1/N)}{\Gamma (1/N)} \,.
\label{kap}
\end{eqnarray}%
Finally, to compare the CFT perturbation result
(\ref{ShortA})-(\ref{ShortB}) with the form factor predictions we
have to compute vacuum expectation values of operators \cite{LZ} in
the conformal normalization (\ref{ConfNorm}).
%
%
%
For $\mu_k$ these important quantities can be derived either from
the reflection relations \cite{FLZZ,Fateev2} or via the deformed
vertex operators \cite{VEV}
\begin{eqnarray}
 \label{MuVev}
&&\langle\mu_{k}\rangle_s= \omega^{sk} \kappa^{2d_{k}}\times\\
&&\ \ \ \exp\int\frac{dt}{t}\left( \frac{\sinh
kut\sinh(N-k)ut}{\sinh t\tanh Nut}-2d_{k}e^{-2t})\right)\,.\nonumber
\end{eqnarray}
Here the index $s = 0,...,N-1$ enumerates the vacuum states in the
$\tilde{Z}_N$ broken phase. (Note, that the correlation functions
$G_\pm$ do not depend on it.)

For the general case,
$\phi_{m,-m}^{k}=(\psi^\dagger)^{l}(\overline{\psi})^{l}\mu_{k}$,
where $m=k-2l$, the vacuum expectation values of fields
are elegantly expressed in terms of $\langle\mu_k \rangle_s$ as %
\[
\frac{\langle\phi_{m,-m}^{k}\rangle_s}
{\langle\mu_{k}\rangle_s}
=\left(\frac{\kappa}{Nu}\right)^{\frac{k^{2}-m^{2}}{2N}}%
{\displaystyle\prod\limits_{i=0}^{l-1}}\bar{\omega}^{2s}
\frac{(i+1)\gamma(\frac{i+1}{N})}{(k-i)\gamma(\frac{k-i}{N})}\,.%
\]
In particular, VEVs of thermal fields $\varepsilon_{k}=(\psi)^{k}%
(\overline{\psi})^{k}\mu_{2k}$ correspond to the $m=0$ case in the
equation above:
\begin{eqnarray}
\langle\varepsilon_{k}\rangle_s=&&\frac{\kappa^{2D_k}}{(Nu)^k}
\frac{(k!)^2}{(2k)!} \left(\frac{\gamma((2k+1)u)}
{\gamma(u)}\right)^{\frac{1}{2}} \times
\\
&&{\displaystyle\prod\limits_{i=1}^{k}}
\frac{\gamma^2(\frac{i}{N})}{\gamma(\frac{2i}{N})\gamma((2i+1)u)}\,.\nonumber%
\end{eqnarray}

Using the method developed in \cite{FFLZZ} we also obtained the
exact result for the expectation values of the normalized descendent
fields $E_{k}=W_{-1}^{(3)}\overline{W}_{-1} ^{(3)}\varepsilon_{k}$
\begin{eqnarray}
\frac{\langle E_{k}\rangle_s}{\langle\varepsilon_{k}\rangle_s}
\,\nonumber
=&& \ \kappa^{2}\frac{(N+2)^{2}}{2N}%
\frac{\Gamma^{2}(1+\frac{k+1}{N})\Gamma^{2}(1-\frac{k}{N})}{\Gamma^{2}%
(1-\frac{k+1}{N})\Gamma^{2}(1+\frac{k}{N})}\times\\
&&\frac{\Gamma(1+\frac{2k}{N})\Gamma(1-\frac{2k+2}{N})}{\Gamma
(2-\frac{2k}{N})\Gamma(2+\frac{2k+2}{N})} \,.%
\end{eqnarray}

Collecting these results we can analyze the short and long distance
behaviors  of correlators. Figs. 1-2 demonstrate typical results
obtained by numerical computations. The lower lines correspond to
the short distance asymptotic from the CFT perturbation theory
(\ref{ShortA})-(\ref{ShortB}) and the upper lines give the form
factor decomposition up to two particle. We see that both expansions
are in a perfect agreement at the intermediate distances and
therefore provide good numerical data for $G_\pm$ at all scales.
This confirms the proposed identifications of form factors and also
gives one of few examples of unitary theories where the correlators
can be effectively studied in the scheme of Ref. \cite{Zam91}.
%
\centerline{\epsfxsize 8.0 truecm \epsfbox{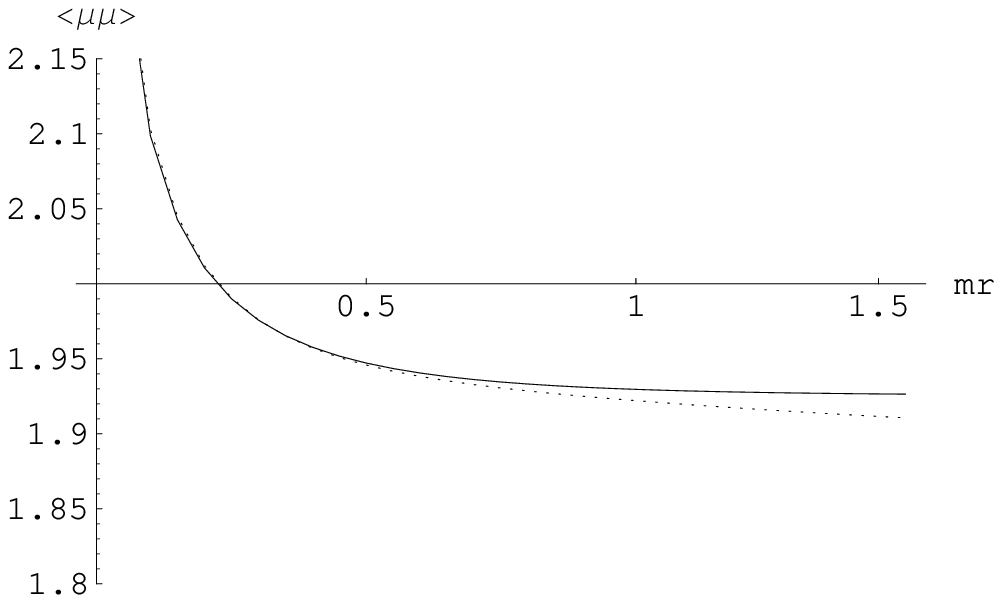}}
\centerline{Fig 1. Disorder correlator $G_-(r)$. N=7 Model.}
\centerline{ }

\centerline{\epsfxsize 8.0 truecm \epsfbox{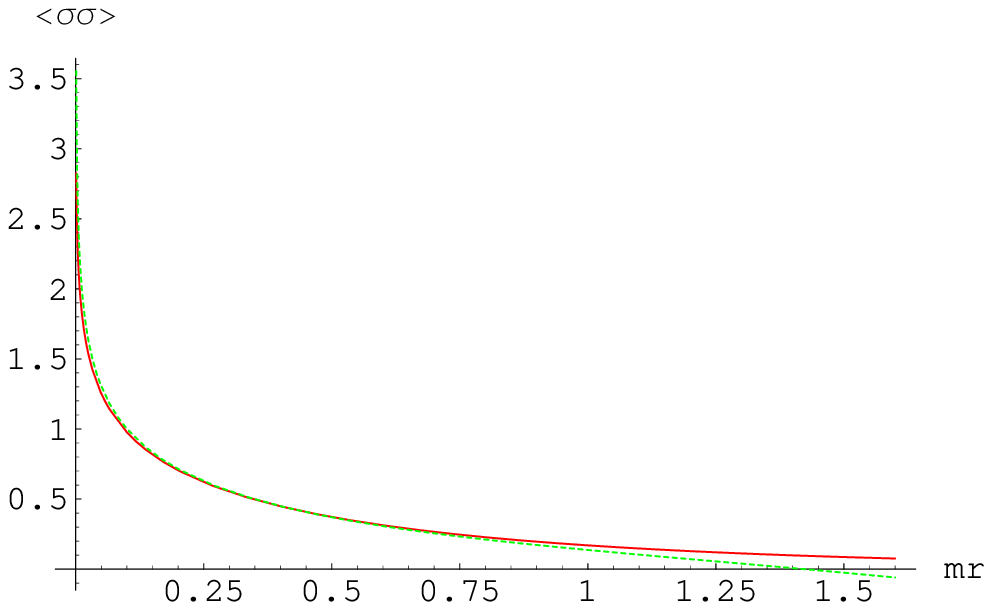}}
\centerline{Fig 2. Order correlator $G_+(r)$. N=7 Model.}
\centerline{ }

%

\vspace{0.2cm} {\it 5. Discussions}
\par\noindent
Algebraic relations among form factors of  $Z_N$ models can be
extended for higher Fourier modes of parafermions as well. Though,
in general, it is not clear how to confirm the proposals, we were
able to provide an additional check supporting the validity of our
construction. Namely, we verified the form factor counterpart of the
quantum equations of motion, which look like:
\begin{eqnarray}
&& \bar{\partial} \psi=(1-D_1)\lambda\ (\psi)\varepsilon_1\,,\cr &&
\partial \bar{\psi}= (1-D_1)\lambda\ (\bar{\psi})\varepsilon_1\,.
\label{PFff}
\end{eqnarray}
The form factors
of operators $(\psi)\varepsilon_1$, $(\bar{\psi})\varepsilon_1$ are
defined as in Eqs. (\ref{PFOne})-(\ref{PFTwo}), while to compute
form factors of currents in the left hand side, we expanded
corresponding matrix elements from the lattice theory \cite{ABFII}
up to first non vanishing terms in the parameter of the lattice
spacing. Equations (\ref{PFff}) explain the relations between form
factors of $\psi$ and $\varepsilon_1$ originally found and discussed
by F. Smirnov \cite{Smir} for the $Z_3$ model.

Notice, that for small number of particles the form factors obtained
from Eqs. (\ref{energy2}), (\ref{PFOne})-(\ref{PFTwo}), (\ref{PFff})
agree up to normalization factors with the results for $Z_3$ case
\cite{Smir} (see also Ref. \cite{Cass}) and for $Z_N$ case
\cite{Bab}.



\vspace{0.2cm} {\it Acknowledgments}

\vspace{0.2cm}

\par\noindent
We would like to thank M. Jimbo, S. Lukyanov, M. Lashkevich and F.
Smirnov for useful discussions. This work was supported by grant
INTAS 03-51-3350, by the Programm of support of leading scientific
schools 2004.2003.2 and by contract EUCLID HRPN-CT-2002-00325. Y.P
was also supported by Russian Science Foundation and by grants
RFBR-04-02-16027, MD-3702.2005.2. V.A.F is grateful to Institute of
Strongly Correlated and Complex Systems at BNL for hospitality.



\vspace{0.2cm} {\it A. Properties of two particle form factors}
\par\noindent
Let the contour $C$ goes from infinity above the real axe, then
around zero and then to infinity below the real axe. Introduce the
notation
\[
S_2(x)=\exp \frac{1}{2}\int_C \frac{dt}{2\pi i
t}\frac{\sinh(x-2\pi)t}{\sinh^2(\pi t)}\log (-t)\,.
\]
Then the functions $\zeta^{(\dagger)}$ appearing in the equation
(\ref{expo1}) read
\begin{eqnarray}
&& \zeta(\beta)=\frac{i\sinh(\frac{\beta}{2})}
{2\sinh(\frac{\beta}{2}+\frac{i\pi}{N})\sinh(\frac{\beta}{2}-\frac{i\pi}{N})}\times
\\ && \hspace{2.5cm} \frac{S_2(i\beta+2\pi
+\frac{2\pi}{N})S_2(-i\beta+\frac{2\pi}{N})}{S_2^2(2\pi
+\frac{2\pi}{N})}\,, \cr
&&\zeta^\dagger(\beta)=\frac{1}{\cosh\frac{\beta}{2} }
\frac{S_2^2(2\pi +\frac{2\pi}{N})} {S_2(i\beta+3\pi
+\frac{2\pi}{N})S_2(-i\beta+\pi+\frac{2\pi}{N})}\,. \nonumber
\end{eqnarray}
These functions satisfy the following important for us properties
\begin{eqnarray} && \frac{\zeta(\beta)}{\zeta(-\beta)}=S(\beta)\,, \quad
\frac{\zeta^\dagger(\beta)}{\zeta^\dagger(-\beta)}=S^\dagger(\beta)\,,\\
%
&& \zeta(\beta)=\zeta(2\pi i -\beta) \,, \quad
\zeta^\dagger(\beta)=\zeta^\dagger(2\pi i -\beta) \,,\cr
%
&& \zeta(\beta)\zeta^\dagger(\beta+i\pi)=\left({ i
\sinh\left(\frac{\beta}{2}-\frac{i\pi }{N}\right)}\right)^{-1}\,,\cr
%
%
%
&&\prod_{j=1}^{N}\zeta\left( \beta+\frac{2j-1}{N}i\pi \right)=\cr
&&\qquad \frac{S_2^{-2N}(2\pi
+\frac{2\pi}{N})}{4\sinh\left(\frac{\beta}{2}+\frac{i\pi}{2N}\right)
\sinh\left(\frac{\beta}{2}-\frac{i\pi}{2N}\right)} \,.\nonumber
\end{eqnarray}
%


\begin{thebibliography}{100}
\bibitem{BPZ}{A.~A.~Belavin, et al.}
              {\em Nucl.\ Phys.} {\bf B241},  333 (1984)

\bibitem{DotsFat}{V.S. Dotsenko, V.A. Fateev.
{\em Nucl. Phys.} {\bf B240}[{\bf  FS12}], 312(1984); {\em Nucl.
Phys.} {\bf B251}[{\bf FS13}], 691 (1985)}


\bibitem{ZaFa85}
A.~B.~Zamolodchikov, V.~A.~Fateev. {\em Sov. Phys. JETP}, {\bf
62}(2), 215 (1985)


\bibitem{Zam}
{A.~B.~Zamolodchikov.  {\em Adv. Stud. in Pure Math.} {\bf 19}, 641
(1989) }

\bibitem{KaWe78}
M.~Karowski, P.~Weisz. {\em Nucl.~Phys.} {\bf B139}, 455 (1978)

\bibitem{smirnovbook}
F.~A.~Smirnov. Form factors in completely integrable models of
quantum field theory. Singapore: World Scientific (1992)

\bibitem{Jim}
J.~L.~Cardy, G.~Mussardo. {\em Nucl. Phys.} {\bf B340}, 387, (1990);
A.~Koubek.  {\em Nucl.Phys.} {\bf B428}, 655 (1994); M.~Jimbo, et
al. [math-ph/0303059];

\bibitem{Del} G.~Delfino, G.~Niccoli. {\em Nucl. Phys.} {\bf B707}, 381
(2005); [hep-th/0501173].


\bibitem{KoSw}
R.~Koberle, J.~A.~Swieca. {\em Phys. Lett.} {\bf B86}, 209 (1979)

\bibitem{Fat91}
V.~A.~Fateev. {} {} {\em Int. J. Mod. Phys.}, {\bf A6}, 2109 (1991)



\bibitem{Luk95}
S.~Lukyanov. {\em Comm. Math. Phys.} {\bf 167}, 183 (1995); {\em
Mod. Phys. Lett.} {\bf A12}, 2543 (1997); {\em Phys.~Lett.}, {\bf
B408}, 192, 1997.


\bibitem{ZF}
S.~Lukyanov, Y.~Pugai. {\em J. Exp. Theor. Phys.} {\bf 82}, 1021
(1996); {\em Nucl. Phys.} {\bf B473}, 631, 1996.


\bibitem{Smir}
F.~A.~Smirnov. {\em Comm.~Math.~Phys.}, {\bf 132}, 415, (1990);
A.~N.~Kirillov, F.~A.~Smirnov. preprint ITF-88-73R, Kiev (1988)


\bibitem{ABFII}
M.~Jimbo, et al. {\em J. Statist. Phys.} {\bf 102}, 883 (2001)

\bibitem{ABF}
G.~E.~Andrews, et al.
{\em J. Stat. Phys.}
{\bf 35}, 193 (1984)

\bibitem{Zam91}
Al.~B.~Zamolodchikov. {\em Nucl.~Phys.}, {\bf B348}, 619 (1991)




\bibitem{VEV}
Y.~Pugai. {\em JETP Lett.} {\bf 79}, 457 (2004)

\bibitem{Fat94}{V.~A.~Fateev. {\em Phys. Lett.} {\bf B324}, 45 (1994)}


\bibitem{LZ}
{S.~Lukyanov, A.~Zamolodchikov. {\em Nucl. Phys.} {\bf B493}, 571
(1997)};
{V.~A.~Fateev, et al. {\it Nucl. Phys.} {\bf B516}, 652
(1998)}

\bibitem{FLZZ}
V.~A.~Fateev. [hep-th/0103014]

\bibitem{Fateev2}
{V.~A.~Fateev. {\em Mod. Phys. Lett.} {\bf A15}, 259 (2000)}


\bibitem{FFLZZ}
{V.~Fateev et al. {\it Nucl.Phys.} {\bf B540}, 587 (1999)}




\bibitem{Cass} M.~Casselle et al. [hep-th/051168]

\bibitem{Bab} H.~Babujian et al. [hep-th/0510062]




















\end{thebibliography}
\end{document}